\documentclass[draftclsnofoot,onecolumn,12pt]{IEEEtran}
\IEEEoverridecommandlockouts
\usepackage{cite}
\usepackage{amsmath,amssymb,amsfonts}
\usepackage{algorithmic}
\usepackage{amsmath} 
\usepackage{graphicx}
\usepackage{textcomp}
\usepackage{xcolor}
\usepackage{amsmath}
\renewcommand{\vec}[1]{\boldsymbol{#1}}
\usepackage{algorithm}
\usepackage{algorithmic}
\usepackage{tcolorbox}
\usepackage{graphicx} 
\usepackage{float} 
\usepackage{stfloats}
\usepackage[section]{placeins}
\usepackage{subfigure}
\usepackage{booktabs}
\usepackage[numbers,sort&compress]{natbib}
\usepackage{amssymb}
\usepackage{setspace}
\usepackage{multirow}
\usepackage{color}
\usepackage{empheq}
\usepackage{hyperref}
\usepackage{pgfplots}
\fontsize{5.0pt}{\baselineskip}\selectfont
\usepackage{epsfig}
\def\BibTeX{{\rm B\kern-.05em{\sc i\kern-.025em b}\kern-.08em
    T\kern-.1667em\lower.7ex\hbox{E}\kern-.125emX}}

\begin{document}
\pgfplotsset{compat=1.14}
\addtolength{\topmargin}{0.04in}
\newlength{\eqnboxwidth}
\setlength{\eqnboxwidth}{\textwidth}
\addtolength{\eqnboxwidth}{-4em}  % some space before equation numbers
\addtolength{\eqnboxwidth}{-.5in} % width of largest equation number

\title{Monte-Carlo Sampling Approach to Model Selection: A Primer}
\author{Petre~Stoica,~\IEEEmembership{Fellow,~IEEE}, Xiaolei~Shang and Yuanbo~Cheng 
\thanks{This work was supported in part by the Swedish Research Council (VR grants 2017-04610  and 2016-06079), in part by the  National Natural Science Foundation of China under Grant 61771442, and in part by Key Research Program of Frontier Sciences of CAS under Grant QYZDY-SSW-JSC035.}
\thanks{P. Stoica is with the Department of Information Technology, Uppsala University, Uppsala SE-751 05, Sweden (e-mail: ps@it.uu.se).}
\thanks{X. Shang and Y. Cheng are  with the Department of Electronic Engineering and Information
Science, University of Science and Technology of China, Hefei 230027, China
(e-mail: xlshang@mail.ustc.edu.cn and cyb967@mail.ustc.edu.cn).} }
\maketitle
\doublespacing
\section*{Introduction and relevance}
Any data modeling exercise has two main components: parameter estimation and model selection. The latter will be the topic of this lecture note.  More concretely we will introduce several Monte-Carlo sampling-based rules for model selection using the maximum \textit{a posteriori} (MAP) approach. Model selection problems are omnipresent in signal processing applications: examples include selecting the order of an autoregressive predictor, the length of the impulse response of a communication channel, the number of source signals impinging on an array of sensors, the order of a polynomial trend, the  number of components of a NMR signal, and so on.
\par We will use the following main notation and definitions in this lecture note (which we prefer to collect in one place so that they can be easily accessed by the reader): \\
\par$\vec{y}\in \mathcal{R}^{N}$=observed  data. 
\begin{equation*}
    \quad\begin{split}
        \lbrace \mathcal{M}_n \rbrace=&\text{the  set  of  models considered  for} \ \vec{y} \ (n=1,\dots,\bar{n} < \infty). \ \text{We do  not assume that any } \\
        &\text{of these models   is an exact description of the system (or mechanism) that generated }\vec{y}, \\
        &\text{but usually one or several of them will  be at least a good approximation of that system.}  
    \end{split}
\end{equation*}
\par$\vec{\theta}_n\in\mathcal{R}^{d_n}$= parameter vector of model $\mathcal{M}_n$.  
\par$p(\cdot)$  denotes a probability density function (pdf). 
\par${\rm prob}(\cdot)$ denotes the probability of a random variable/event.
\par$\hat{\vec{\theta}}_n=\arg \max_{\vec{\theta}_n}p(\vec{y} | \mathcal{M}_n,\vec{\theta}_n)$= the  maximum likelihood estimate (MLE) of $\vec{\theta}_n$. 
\begin{equation*}
    \quad \begin{split}
        \vec{\theta}_n^{*}=&\lim_{N\to \infty}\hat{\vec{\theta}}_n \ \text{(which usually exists under weak conditions). By the  consistency property of MLE,} \\
        & \text{$\vec{\theta}_n^{*}$ is the true parameter vector when $\mathcal{M}_n$ contains the data generating mechanism.} 
    \end{split} \nonumber
\end{equation*}
\par $\delta(x)=$ Dirac delta (unit impulse at  $x=0$).
\par $ \delta_{i,j}$=Kronecker delta (=1 if $i=j$ and 0 otherwise).
\par AIC=Akaike information criterion. 
\par BIC=Bayesian information criterion. 
\par MAP=Maximum \textit{a posteriori}.\\
Essentially what we will discuss in this lecture note is how to use the MAP approach along with $\vec{y}$ and $\lbrace \hat{\vec{\theta}}_n \rbrace$ to select a model $\mathcal{M}_{\hat{n}}$ out of the set $\lbrace \mathcal{M}_n \rbrace$.  When the models $\lbrace \mathcal{M}_n \rbrace$ are completely specified (i.e. $\vec{\theta}_n$ in each of them is known), the MAP rule consists of choosing the model that maximizes the following conditional/posterior pdf:
\begin{equation}
    \max_{n} \ p(\mathcal{M}_n | \vec{y}) \label{eq:1}
\end{equation}
Because 
\begin{equation}
    p(\mathcal{M}_n | \vec{y})=p( \vec{y}| \mathcal{M}_n)p(\mathcal{M}_n)/p(\vec{y})
\end{equation}
and usually $p(\mathcal{M}_n)=1/ \bar{n}$ (unless we know from apriori information that some models are more preferable than others) we have:
\begin{equation}
    p(\mathcal{M}_n | \vec{y}) \sim p(\vec{y} | \mathcal{M}_n)
\end{equation}
and hence \eqref{eq:1} becomes
\begin{equation}
    \max_{n} \ p(\vec{y}|\mathcal{M}_n) \label{eq:4}
\end{equation}
(see, e.g., \cite{kay1993fundamentals}\cite{stoica2004model}). The above MAP rule maximizes the total (or average) probability of correct selection (or detection) (see the cited references) and therefore it is optimal. The problem is that in most applications it cannot be used as it stands because,  while the form/structure of the models   $\lbrace \mathcal{M}_n \rbrace$ is known, the parameter vectors $\lbrace\vec{\theta}_n\rbrace$ are unknown. In fact, optimal selection rules exist only in very special cases (see, for example, the discussion on p356 in \cite{stoica2004cross}). Therefore in most applications we should be content with using a sub-optimal selection rule that hopefully has good performance. 
\par There are a large number of methods that  are trying to bypass the difficulty caused by the fact that $\lbrace \mathcal{M}_n \rbrace$ are not completely specified. Using $p(\vec{y} | \mathcal{M}_n,\hat{\vec{\theta}}_n)$ instead of $p(\vec{y} | \mathcal{M}_n)$ in \eqref{eq:4} is not a good idea: in particular, for nested models (for which $\mathcal{M}_{n_1} \subset  \mathcal{M}_{n_2}$ if $d_{n_1} < d_{n_2}$), the sequence of likelihoods $\lbrace p(\vec{y} | \mathcal{M}_n,\hat{\vec{\theta}}_n) \rbrace$ increases with $d_n$ and hence the  most complex model (which has the largest number of parameters) will always be selected. If $\lbrace \vec{\theta}_n^{*} \rbrace$ were known, a much better choice would be to compare
\begin{equation}
    p(\vec{y} | \mathcal{M}_n, \vec{\theta}_n^{*}), \ n=1,\dots,\bar{n}
\end{equation}
but $\lbrace \vec{\theta}_n^{*} \rbrace$ are almost never known in practical applications. 
\par A successful class of rules is based on penalising the model complexity (basically, its number of parameters $\lbrace d_n \rbrace$) by adding a penalty term to the negative log-likelihood:
\begin{equation}
    -2\ln p(\vec{y} | \mathcal{M}_n,\hat{\vec{\theta}}_n) +\gamma d_n \label{eq:6}
\end{equation}
Different values of $\gamma$ are obtained from different types of statistical or information theory considerations: for example, $\gamma=2$ for AIC and $\gamma=\ln N$ for BIC. AIC is minimax optimal and BIC is consistent, but as alluded in the discussion of \cite{stoica2004cross} mentioned above neither is  optimal (in the sense of maximizing the probability of selection) and therefore they both can  in principle be outperformed by other rules.
\par \textit{Remark}: The form of  the BIC rule in \eqref{eq:6} (with $\gamma=\ln{N}$) is the one most commonly used in applications, but it is (asymptotically) valid only in the cases in which the variances of the estimation errors of elements of $\hat{\vec{\theta}}_n$ all go to zero  as $1/N$, as $N\to \infty$.  If the variances go to zero at different rates, which is the case for some models, then this fact should be taken into account as it leads  to a different penalty term for BIC than the one in \eqref{eq:6} (see e.g. \cite{stoica2004model} and \cite{stoica2012proper} for details on this aspect).
\par In an attempt to find better rules researchers have tried to approximate $p(\vec{y} | \mathcal{M}_n)$, when $\{\vec{\theta}_n\}$ are unknown, using a prior pdf $p(\vec{\theta}_n | \mathcal{M}_n)$ for the parameter vector:
\begin{equation}
    p(\vec{y} | \mathcal{M}_n)=\int_{D_n} p(\vec{y} | \mathcal{M}_n, \vec{\theta}_n) p(\vec{\theta}_n | \mathcal{M}_n) d\vec{\theta}_n \label{eq:7}
\end{equation}
where $D_n \subset \mathcal{R}^{d_n}$ denotes the support domain. To make use of \eqref{eq:7} we need to solve two main problems:
\begin{enumerate}
\item[A)] Select  $p(\vec{\theta}_n | \mathcal{M}_n)$ and implicitly $D_n$
\item[B)] Evaluate the integral in \eqref{eq:7}
\end{enumerate}
\par  Concerning  A, in some applications, there is information about  $\vec{\theta}_n$ that allows the formulation of a prior pdf. When such information does exist, it must be used: for example this information could emphasize a subset of $D_n$ on which $p(\vec{y} | \mathcal{M}_n, \vec{\theta}_n)$ takes on comparatively small values - the corresponding choice of $p(\vec{\theta}_n | \mathcal{M}_n)$ would in such a case avoid  the data overfitting associated with too complex models (see, e.g., \cite{llorente2020marginal}). However this type of prior information is rarely available in signal processing applications in which most often the only information that we have about  the model parameters is obtained aposteriori after processing the data to get $\lbrace \hat{\vec{\theta}}_n \rbrace$. This is the case we will consider here: we will choose both $D_n$ and $p(\vec{\theta}_n|\mathcal{M}_n)$ based on $\hat{\vec{\theta}}_n$ and its properties, see the next section. Because $p(\vec{\theta}_n|\mathcal{M}_n)$ then will depend on the data, the corresponding MAP approach will be an empirical Bayesian methodology.
\par Regarding  B, there is a huge number of methods that can be employed to evaluate integrals  of the above type (see \eqref{eq:7}) using direct basic sampling, nested sampling, bridge sampling, path sampling,  layered importance sampling, and so forth (see, e.g., \cite{llorente2020marginal}). We will consider direct sampling, either from the prior pdf or from an importance (also called proposal) pdf, with the goal of assessing what such simple sampling methods have to offer. Our main focus will  be on A rather than on B. By keeping the discussion on B as simple as possible our hope is that more readers will understand the basics of Monte-Carlo sampling and will feel motivated to deepen  their knowledge beyond the direct sampling methods discussed in this lecture note. 
\section*{Prerequisites}
While we will do our best to make this lecture note as self-contained as possible, basic knowledge of statistical signal processing  as well as estimation and detection theory  will be beneficial for fully understanding it.

\section*{Model Selection rules}
The main idea of all the rules presented in what follows is simple: in a nutshell we will try to get an estimate $\hat{p}(\vec{y} | \mathcal{M}_n)$ of the posterior pdf in \eqref{eq:7} and then choose the model that maximizes it:
\begin{equation}
    \hat{n}=\arg \max_{n} \ \hat{p}(\vec{y} | \mathcal{M}_n)
\end{equation}
Let 
\begin{equation}
    \tilde{\vec{\theta}}_n=\arg \min_{\vec{\theta}_n} p(\vec{y} | \mathcal{M}_n,\vec{\theta}_n) \nonumber
\end{equation}
Then we clearly have (see \eqref{eq:7}):
\begin{equation}
    p(\vec{y} | \mathcal{M}_n,\tilde{\vec{\theta}}_n) \leq p(\vec{y} | \mathcal{M}_n) \leq p(\vec{y} | \mathcal{M}_n, \hat{\vec{\theta}}_n)  \label{eq:9}
\end{equation}
The lower and upper bounds in \eqref{eq:9} correspond to the following priors: $\delta(\vec{\theta}_n-\tilde{\vec{\theta}}_n)$ and $\delta(\vec{\theta}-\hat{\vec{\theta}}_n)$, respectively, with the latter inducing no penalization at all on  the likelihood $p(\vec{y} | \mathcal{M}_n, \hat{\vec{\theta}}_n)$ and the former causing the maximum possible penalization. Any other choice of $p(\vec{\theta}_n | \mathcal{M}_n)$ will penalize the model's complexity in between these extremes. The choice of the prior thus is quite important as it can significantly affect the final result of the model choice based on \eqref{eq:7} (see, e.g.,  \cite{llorente2020marginal} for more details on this aspect).
\par Consider the following  ``concentration ellipsoid'':
\begin{align}
    C_n=\lbrace \vec{\theta}_n \ | \ (\vec{\theta}_n - \hat{\vec{\theta}}_n)^T\vec{J}_n(\vec{\theta}_n-\hat{\vec{\theta}}_n) \leq {\mu}_n \rbrace \label{eq:10}
\end{align}
where $\vec{J}_n$ is the sample Fisher information matrix (FIM) associated with $\mathcal{M}_n$:
\begin{align}
    \vec{J}_n = -\frac{\partial^2 \ln p(\vec{y} | \mathcal{M}_n,\vec{\theta}_n)}{\partial \vec{\theta}_n \partial \vec{\theta}_n^T} \left. \right|_{\vec{\theta}_n=\hat{\vec{\theta}}_n}
\end{align}
(assumed to be nonsingular) and 
\begin{equation}
    \mu_n=6+2d_n \label{eq:12}
\end{equation}
(other choices of $\mu_n$ are also possible, but the above one worked well in our numerical experiments). When $\mathcal{M}_n$ is a good approximation of the data generating mechanism, $\vec{\theta}_n^{*}$ will belong to $C_n$ with a probability of about  0.99. This follows from the fact that in such a case
\begin{align}
    (\hat{\vec{\theta}}_n - \vec{\theta}_n^{*}) \sim \mathcal{N}(\vec{0},\vec{J}_n^{-1})
\end{align}
(see, e.g., \cite{soderstrom1989system}\cite{kay1998fundamentals}) and hence 
\begin{equation}
    (\hat{\vec{\theta}}_n- \vec{\theta}_n^{*})^T \vec{J}_n (\hat{\vec{\theta}}_n- \vec{\theta}_n^{*})  \sim \chi^2(d_n)
\end{equation}
The probability of $\vec{\theta}^{*}_n \in C_n$ can then be obtained from tables of the $\chi^{2}$ distribution (viewing $\vec{\theta}_n^{*}$ as a ``random variable'' conditioned on $\hat{\vec{\theta}}_n$, which is given): for the choice of $\mu_n$ in \eqref{eq:12} this probability varies from 0.995 for $d_n=1$ to 0.996 for $d_n=10$. 

\textit{Remark}: When $\mathcal{M}_n$ is a poor approximation of the system that generated $\vec{y}$, in principle $\vec{J}_n$ in \eqref{eq:10} should be replaced with the matrix given by the so-called ``sandwich  formula'' (\cite{soderstrom1989system}\cite{white1996estimation}). However, to keep the discussion here as simple as possible we will use $\vec{J}_n$ in what follows.

\par The prior pdf's we will  consider are presented in the following subsections.
\subsection*{UE (Uniform prior, Ellipsoid set)}
We let $\vec{\theta}_n$ be uniformly distributed on $C_n$:
\begin{equation}
    p(\vec{\theta}_n | \mathcal{M}_n)=  \left\{\begin{aligned}
        1/V(C_n) \quad  &{\rm  if} \quad \vec{\theta} \in C_n \\
        \! 0  \quad & {\rm else} \\ 
\end{aligned}
\right.  \label{eq:15}
\end{equation}
For later use we note that the volume of $C_n$ is given by
\begin{align}
    V(C_n) = (\mu_n)^{d_n/2} V_{d_n}/ |\vec{J}_n|^{1/2}
\end{align}
where $V_{d_n}$ is the volume of the unit ball in $d_n$ dimensions (which can be easily computed recursively in $d_n$). 

\par Let $\lbrace \vec{\theta}_{n}^m \rbrace$ (for $m=1,\dots,M_n$) denote $M_n$ samples drawn from the uniform pdf in \eqref{eq:15}. We can generate $\lbrace \vec{\theta}_n^m \rbrace$ in the box $B_n$ defined in Lemma 1 in the Appendix (see equation \eqref{eq:box} there) and retain the vectors that lie in $C_n$: according to Lemma 2 in the Appendix the so-obtained $\lbrace \vec{\theta}_n^m \rbrace$ have a uniform distribution on $C_n$. Using $\lbrace \vec{\theta}_n^m \rbrace$ we can obtain an unbiased estimate of the posterior pdf in \eqref{eq:7} as follows:
\begin{equation}
\framebox{\parbox{\eqnboxwidth}{\centerline{$ \displaystyle
       \hat{p}_{\rm UE}(\vec{y}| \mathcal{M}_n) =\frac{1}{M_n}\sum_{m=1}^{M_n} p(\vec{y} | \mathcal{M}_n,\vec{\theta}_n^m) 
$ }
}}
\label{eq:17}
\end{equation} 

\par It is our experience that a reasonably small value of $M_n$, such as $M_n=1000$, yields accurate results in many cases, and therefore choosing a much larger value for $M_n$ is often unnecessary (however see the discussion in the next paragraph). 
\par While \eqref{eq:17} is appealing because it is one of the most direct sampling-based unbiased estimates of $p(\vec{y}| \mathcal{M}_n)$, its variance may be large especially if the variation of $p(\vec{y}| \mathcal{M}_n,\vec{\theta}_n)$ over $C_n$ is significant (see the section titled ``Variance analysis of Monte-Carlo sampling'' in the Appendix; in such a case choosing a larger value for $M$ than $M=1000$ may be justified). A potentially more accurate unbiased estimate of $p(\vec{y}| \mathcal{M}_n)$ can be obtained using importance sampling as described next. 
\subsection*{UEG (Uniform prior, Ellipsoid set, Gaussian proposal)}
As implied by the discussion above, the variance of \eqref{eq:17} would be comparatively small if the pdf from which $\lbrace \vec{\theta}_n^{m} \rbrace$ are drawn was proportional to $p(\vec{y} | \mathcal{M}_n,\vec{\theta}_n)$ over $C_n$ (see, e.g. \cite{mcbook}\cite{robert2013monte}). Here we derive a proposal pdf (a term used in importance sampling, see the cited work) whose variation, at least locally around $\hat{\vec{\theta}}_n$, mimics the variation  of $p(\vec{y}| \mathcal{M}_n,\vec{\theta}_n)$. \\
\noindent First note that (for $N\gg 1$),
\begin{equation}
    \ln p(\vec{y}| \mathcal{M}_n,\vec{\theta}_n) \thickapprox \ln p(\vec{y} | \mathcal{M}_n, \hat{\vec{\theta}}_n) -\frac{1}{2} (\vec{\theta}_n -\hat{\vec{\theta}}_n)^T \vec{J}_n (\vec{\theta}-\hat{\vec{\theta}}_n)  
\end{equation}
and therefore 
\begin{equation}
    p(\vec{y} | \mathcal{M}_n, \vec{\theta}_n) \thickapprox p(\vec{y} | \mathcal{M}_n,\hat{\vec{\theta}}_n) e^{-\frac{1}{2} (\vec{\theta}_n -\hat{\vec{\theta}}_n)^T \vec{J}_n (\vec{\theta}-\hat{\vec{\theta}}_n) } \label{eq:19}
\end{equation}
The following Gaussian pdf,
\begin{equation}
    g(\vec{\theta}_n) = \frac{|\vec{J}_n|^{1/2}}{(2\pi)^{d_n/2}} e^{-\frac{1}{2} (\vec{\theta}_n -\hat{\vec{\theta}}_n)^T \vec{J}_n (\vec{\theta}-\hat{\vec{\theta}}_n) } \label{eq:20}
\end{equation}
is proportional to the (local) approximation of $p(\vec{y} |\mathcal{M}_n,\vec{\theta}_n)$ in \eqref{eq:19}. We will use the above $g(\vec{\theta}_n)$ to estimate \eqref{eq:17}, with the uniform prior pdf in \eqref{eq:15}, employing the importance sampling approach. This approach is based on the simple observation that, for $p(\vec{\theta}_n|\mathcal{M}_n)$ in \eqref{eq:15},
\begin{equation}
    \int_{D_n} p(\vec{y} | \mathcal{M}_n,\vec{\theta}_n) p(\vec{\theta}_n | \mathcal{M}_n) d\vec{\theta}_n = \frac{1}{V(C_n)} \int_{C_n} \frac{p(\vec{y} | \mathcal{M}_n,\vec{\theta}_n)}{g(\vec{\theta}_n)} g(\vec{\theta}_n) d\vec{\theta}_n 
\end{equation}
It follows from the above equation that an unbiased estimate of $p(\vec{y} | \mathcal{M}_n)$ is given by:
\begin{equation}
\framebox{\parbox{\eqnboxwidth}{\centerline{$ \displaystyle
       \hat{p}_{\rm UEG} (\vec{y} | \mathcal{M}_n) =\frac{\rho_n}{M_nV(C_n)} \sum_{m=1}^{M_n} \frac{p(\vec{y}| \mathcal{M}_n,\vec{\theta}_n^m)}{g(\vec{\theta}_n^m)}
$ }
}}
\label{eq:22}
\end{equation} 
where $\lbrace \vec{\theta}_n^m \rbrace$ are drawn from \eqref{eq:20} and used in \eqref{eq:22} only if they belong to $C_n$: $\vec{\theta}_n^m \in C_n$. The factor $\rho_n$ in \eqref{eq:22} is such that the  truncated Gaussian function
\begin{equation}
    \left\{\begin{aligned}
        \frac{1}{\rho_n} g(\vec{\theta}_n) \quad  &{\rm  if} \quad \vec{\theta}_n \in C_n \\
        \! 0  \quad & {\rm else} \\ 
\end{aligned}
\right.  \label{eq:23}
\end{equation}
is a valid pdf. In other words, $\rho_n$ is a normalizing factor given by 
\begin{equation}
    \rho_n=\int_{C_n} g(\vec{\theta}_n) d\vec{\theta}_n
\end{equation}
Note that $\rho_n$ is the probability of a Gaussian random variable drawn from \eqref{eq:20} to belong to $C_n$, or equivalently, the probability of  a $\chi^{2}(d_n)$ random variable to be less than $\mu_n$. The latter probability, and hence $\rho_n$, can be obtained from tables of the $\chi^{2}$ distribution (as already indicated above). Also note that the parameter vectors $\lbrace \vec{\theta}_n^m \rbrace$, generated as explained above, are distributed according to \eqref{eq:23} as they should, which follows from the acceptance-rejection theorem (see e.g. \cite{liu2001monte} and also the section titled ``The acceptance and rejection algorithm'' in the Appendix). 

\par We should stress the fact that the UEG approach estimates the same quantity as the UE approach (i.e. \eqref{eq:7} with the same uniform prior pdf). But, as indicated above, UEG can be expected to be more accurate. In the next subsection we will describe a different choice of the prior pdf, inspired by the $g(\vec{\theta}_n)$ in \eqref{eq:20}. We note in passing that the said $g(\vec{\theta}_n)$ is the asymptotic distribution of the maximum likelihood estimate $\hat{\vec{\theta}}_n$ (under the assumption that $\vec{\theta}_n$ is the true parameter vector). 

\subsection*{GE (Gaussian prior, Ellipsoid set)}
Using  \eqref{eq:23} as the prior pdf leads to the following unbiased estimate of the $p(\vec{y} | \mathcal{M}_n)$ in \eqref{eq:7}:
\begin{equation}
\framebox{\parbox{\eqnboxwidth}{\centerline{$ \displaystyle
     \hat{p}_{\rm GE} (\vec{y} | \mathcal{M}_n) = \frac{1}{M_n} \sum_{m=1}^{M_n} p(\vec{y} | \mathcal{M}_n,\vec{\theta}_n^m)
$ }
}}
\label{eq:25}
\end{equation} 
where $\lbrace \vec{\theta}_n^m \rbrace$ are generated exactly as in UEG. As explained above by virtue of the acceptance-rejection theorem, the distribution of $\lbrace \vec{\theta}_{n}^m  \rbrace$ is given by \eqref{eq:23} as desired.
\par Note that GE penalizes the model complexity less than UE and UEG because parameter vectors in \eqref{eq:7}, which are far from $\hat{\vec{\theta}}_n$, receive a smaller weight in GE than in the other two methods. Also, it is interesting to observe that the estimates in \eqref{eq:22} and \eqref{eq:25} have rather different expressions in spite of the fact that the parameter samples $\lbrace \vec{\theta}_n^m \rbrace$ are drawn from the same distribution in both equations (the reason for this significantly different expressions is that \eqref{eq:22} is based on an uniform prior whereas \eqref{eq:25} uses a truncated  Gaussian prior).
\subsection*{UB (Uniform prior, Box set)}
In the process of generating the vectors $\lbrace \vec{\theta}_n^m \rbrace$ for UE we discarded any sample that did not belong to $C_n$. To avoid wasting those samples, we can extend $C_n$ to the box $B_n$ (see \eqref{eq:box} in  the Appendix) and consequently consider the following uniform distribution as the prior pdf:
\begin{equation}
    p(\vec{\theta}_n | \mathcal{M}_n)=\left\{\begin{aligned}
        \frac{1}{V(B_n)}  \quad  &{\rm  if} \quad \vec{\theta}_n \in B_n \\
        \! 0  \quad & {\rm else} \\ 
\end{aligned}
\right.  \label{eq:26}
\end{equation}
Using \eqref{eq:26} in \eqref{eq:7} leads to the following unbiased estimate of $p(\vec{y}| \mathcal{M}_n)$:
\begin{equation}
\framebox{\parbox{\eqnboxwidth}{\centerline{$ \displaystyle
       \hat{p}_{\rm UB} (\vec{y} | \mathcal{M}_n) = \frac{1}{M_n} \sum_{m=1}^{M_n} p (\vec{y} | \mathcal{M}_n,\vec{\theta}_n^m)
$ }
}}
\end{equation} 
where $\lbrace  \vec{\theta}_n^m \rbrace$ are drawn from \eqref{eq:26}. Because it uses a larger set than $C_n$, UB penalizes the model complexity more heavily than both UE and UEG as well as GE. 
\par Finally we remark on the fact that all four methods presented above require the evaluation of the likelihood function $p(\vec{y} | \mathcal{M}_n, \vec{\theta}_n)$, which can take on rather small values (especially for $N\gg 1$). To avoid possible numerical problems we can evaluate this function as: 
\begin{equation}
    p(\vec{y} | \mathcal{M}_n,\vec{\theta}_n)=e^{\ln p(\vec{y} | \mathcal{M}_n, \hat{\vec{\theta}}_n)} e^{\ln p(\vec{y}| \mathcal{M}_n,\vec{\theta}_n)-\ln p(\vec{y}| \mathcal{M}_n,\hat{\vec{\theta}}_n)}
\end{equation}
or use symbolic computations in MATLAB. 
\section*{Numerical performance study}
As  mentioned in the previous section, GE penalizes model complexity the least and UB the most with UE  (and its close relative UEG) in between. We have found out empirically that UE, UEG and GE have a tendency to select non-parsimonious models in many cases,  a fact which renders them less competitive than UB. Consequently, in this section, we will focus on the UB rule whose performance for model selection will be compared to that of AIC and BIC. 

We will consider polynomial models described by the equation:
\begin{equation}
    \mathcal{M}_n: y(t)=\sum_{i=1}^n a_i x_i(t) +e(t) \quad t=1,\dots,N \label{eq:polynomial}
\end{equation}
where 
\begin{equation}
    x_i(t)=\left[ -5 + 10(t-1)/(N-1) \right]^{i-1}
\end{equation}
and $e(t)$ is assumed to be a Gaussian white noise with zero mean and variance denoted $\sigma^2$. The data are generated with \eqref{eq:polynomial} for $n=4$, $\sigma^2=1$ and $a_1,a_2=0.1$, $a_3=-0.3$ and $a_4=0.4$.

For the linear regression model in \eqref{eq:polynomial}, the likelihood function has the following simple expression (here $\vec{y}=\begin{bmatrix}y(1),\dots,y(N) \end{bmatrix}^T$):
\begin{equation}
    p(\vec{y}|\mathcal{M}_n,\vec{\theta}_n)=\frac{1}{(2\pi)^{N/2}\sigma^N} e^{-\frac{1}{2\sigma^2}\sum_{t=1}^N \left[y(t)-\vec{\varphi}^T(t)\vec{\theta}_n\right]^2} \nonumber
\end{equation}
where 
\begin{align}
    \vec{\theta}_n&=\begin{bmatrix}a_1, \dots, a_n \end{bmatrix}^T  \nonumber \\
    \vec{\varphi}(t)&=\begin{bmatrix}x_1(t),\dots, x_n(t) \end{bmatrix}^T \nonumber
\end{align}
Furthermore, the MLE of $\vec{\theta}_n$ and the corresponding FIM are given by:
\begin{align}
    \hat{\vec{\theta}}_n & = \left[ \sum_{t=1}^N \vec{\varphi}(t)\vec{\varphi}^T(t)\right]^{-1}\cdot \left[ \sum_{t=1}^N \vec{\varphi}(t)y(t)\right] \nonumber \\
    \vec{J}_n &= \frac{1}{\sigma^2} \left[\sum_{t=1}^N \vec{\varphi}(t)\vec{\varphi}^T(t)\right]
\end{align}
(we assume for the sake of simplicity  that $\sigma^2$ is known, see e.g. \cite{soderstrom1989system}\cite{kay1998fundamentals}). We use 1000 replications of the noise sequence to estimate the frequencies of  the selected orders $\hat{n}\in [1,6]$ by AIC, BIC and UB (with $M=1000$).
\begin{figure}[htb]
\centering
\begin{minipage}[t]{0.38\linewidth}
\centering
\centerline{\epsfig{figure=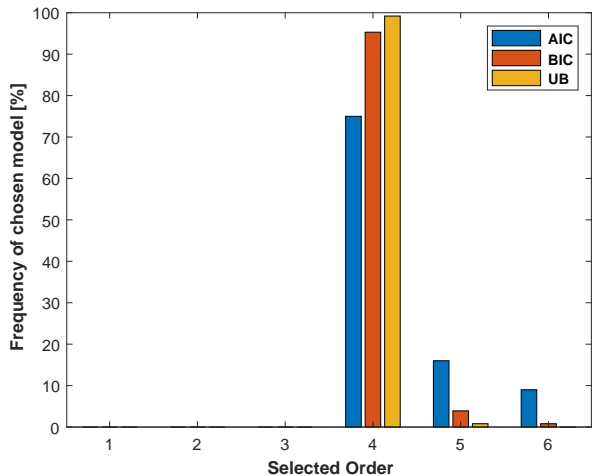,width=9cm}}
\end{minipage}
\caption{Frequencies of selected orders $\hat{n}\in \left[1,6\right]$ for $N=100$.}
\label{fig:fig1}
\end{figure}
Figure \ref{fig:fig1} shows the histograms of the orders selected by these three rules. In this case UB is slightly better than BIC, which is better than AIC.
\begin{figure}[htb]
\centering
\begin{minipage}[t]{0.38\linewidth}
\centering
\centerline{\epsfig{figure=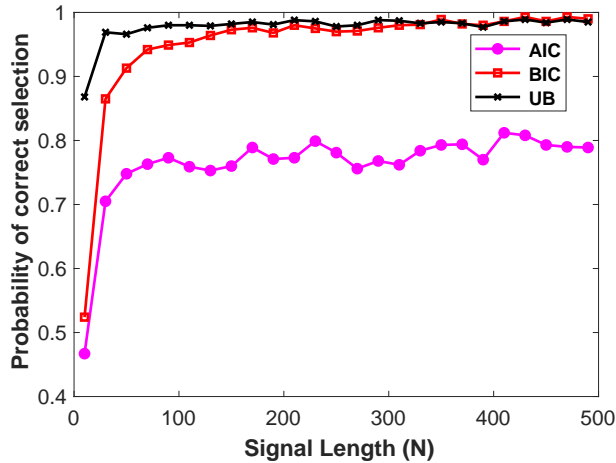,width=9cm}}
\end{minipage}
\caption{Estimated probability of correct selection versus $N$.}
\label{fig:fig2}
\end{figure}

In Fig. \ref{fig:fig2} we show the empirical probabilities of correct selection (i.e. $\hat{n}=4$) versus the number of data samples, $N$. The ranking of the three rules is similar to that in Fig. \ref{fig:fig1}. In the case of BIC and UB the probability of correct selection approaches 1 as $N$ increases, whereas for AIC this probability reaches a ceiling of about 0.8 (it is a well-known fact that asymptotically AIC has a probability of correct selection only slightly larger than 0.8).

For obvious reasons using sequences given by a single data generating mechanism for comparing model selection rules can lead to biased conclusions.  A way of preventing this from happening in the current case is to average the results over many polynomials. This is exactly what we will do in the remaining part of this section. We now use \eqref{eq:polynomial} to generate data sequences $\vec{y}$ of length $N$, for orders $n=1,\dots,6$,  $m=100$ polynomial coefficient vectors $\vec{\theta}_n$ uniformly drawn from the cube $\left[-0.5,  0.5\right]^n$  and  $r=1000$  noise realizations. In this way, we will obtain $6\cdot 10^5$ data sets from which we can calculate the \textit{average} probability of selection as follows:
\begin{equation}
    \frac{1}{6\cdot10^5} \sum_{n=1}^6\sum_{m=1}^{100}\sum_{r=1}^{1000}\delta_{n,\hat{n}_{n,m,r}} \label{eq:101}
\end{equation}
where $\hat{n}$ denotes the selected order (which is a function of $n$, $m$ and $r$).
In Fig. 3 we plot \eqref{eq:101} versus $N$. 
\begin{figure}[htb]
\centering
\begin{minipage}[t]{0.38\linewidth}
\centering
\centerline{\epsfig{figure=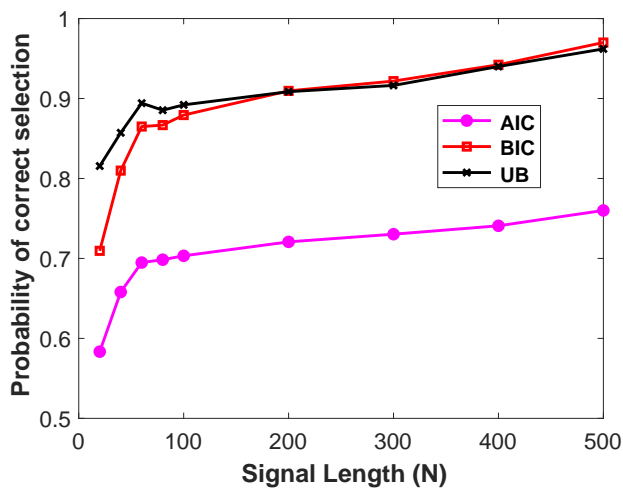,width=9cm}}
\end{minipage}
\caption{Estimated average probability of correct selection versus $N$.}
\label{fig:fig3}
\end{figure}
The advantage of UB over BIC (for $N$ up to about 150) and especially AIC (for all values of $N$) is similar to what we observed in Fig. \ref{fig:fig2}. However, the convergence to asymptotics (as $N$ increases) is slower than in Fig. \ref{fig:fig2}. This is likely due to a higher probability of underestimating the true order than in Fig. \ref{fig:fig2}: indeed in the present case some of the generated coefficients $a_n$ can be rather close to zero and therefore the $n$-th order polynomial that was used to generate the data can be well approximated by an $(n-1)$-th order polynomial. We also note that in the case of Fig. \ref{fig:fig3} we had to use a larger $M$ than in Fig. \ref{fig:fig2}, as $M=1000$ (used to obtain Figs. \ref{fig:fig1} and \ref{fig:fig2}) turned out to be too small for the runs in Fig. \ref{fig:fig3} with polynomials of order 5 and 6. 

\section*{What we have learned}
AIC and BIC are the rules most commonly used in the data modeling applications in which the parameters are estimated via the maximum  likelihood method. These workhorses of model selection have several distinct advantages: they are  easy to understand, easy to code and require a negligible amount of additional computing time. On the other hand, their theoretical foundations and derivations from statistical or information theory principles are somewhat  complicated, but that aspect is not of paramount importance to the practitioners who have used these rules in a myriad of applications and have noticed their satisfactory performance. There have been many attempts to propose alternative rules, some of them quite ingenious, but none has received the kind of significant attention that AIC and BIC have.
 
The Monte-Carlo sampling approach to model selection discussed in this lecture note is also easy to understand and easy to code, but it requires longer computation times than AIC and BIC do. However,   nowadays the mentioned difference in computation times is no longer a decisive factor. Using the UB rule as an example, while its running time  can be one or even two orders of magnitude longer than that of BIC or AIC, its use on a standard laptop required less than 1 sec in our numerical experiments with $M=1000$ and $N=1000$, which makes it perfectly usable in many applications. From a theoretical standpoint, the derivation of UB (or the other MAP-type rules discussed in this lecture note) is quite simple, which is a potential advantage over AIC and BIC. However, the most important advantage of UB lies in the fact that it can perform better than AIC and BIC, or at least not significantly worse than  the best of the latter two rules.   

A goal of this lecture note was to demystify the Monte-Carlo sampling approach to model selection, for those readers less familiar with it, which in its basic form is quite simple as it was shown in the previous sections. Another goal was to discuss the main choices a user of this approach has to make. These choices concern  the prior distribution and its support domain, and they are important as they can make or break a model selection rule. Our hope is that the readers of this lecture note will feel motivated to deepen their knowledge on the subject matter and also get new ideas for model selection rules  with enhanced performance. 
\appendix
\textbf{Lemma 1.} The smallest box that encompasses $C$ in \eqref{eq:10} is given by
\begin{equation}
    B=\prod_{k=1}^{d}\left[ \hat{\theta}_k - \sqrt{\mu(\vec{J}^{-1})_{kk}}, \ \hat{\theta}_k + \sqrt{\mu(\vec{J}^{-1})_{kk}} \right] \label{eq:box}
\end{equation}
where $\hat{\theta}_k$ is the $k$-th component of $\hat{\vec{\theta}}$, and $(\vec{J}^{-1})_{kk}$ is the $k$-th entry on the diagonal of $\vec{J}^{-1}$ (note that here we omit the sub-index $n$ of the different variables to simplify the notation). \\
\textit{Proof.} Let 
\begin{equation}
    \vec{e}_k=\begin{bmatrix}\underbrace{ 0, \dots, 0, 1}_{k}, 0, \dots, 0 \end{bmatrix}^T
\end{equation}
We want to determine the points on the surface of $C$ that are at maximum distance from $\hat{\vec{\theta}}$ along the axes:
\begin{align}
    \max_{\vec{\theta}} |\vec{e}_k^T(\vec{\theta}-\hat{\vec{\theta}})| \quad {\rm s.t.} \ (\vec{\theta}-\hat{\vec{\theta}})^T\vec{J}(\vec{\theta}-\hat{\vec{\theta}}) = \mu, \  (k=1,\dots,d) 
\end{align}
A simple use of the Cauchy–Schwarz inequality yields:
\begin{align}
    |\vec{e}_k^T(\vec{\theta}-\hat{\vec{\theta}})|^2=|\vec{e}_k^T\vec{J}^{-1/2}\vec{J}^{1/2}(\vec{\theta}-\hat{\vec{\theta}})|^2  \leq (\vec{e}_k^T\vec{J}^{-1}\vec{e}_k)(\vec{\theta}-\hat{\vec{\theta}})^T\vec{J}(\vec{\theta}-\hat{\vec{\theta}}) =\mu (\vec{J}^{-1})_{kk}
\end{align}
where the equality holds if
\begin{equation}
  \theta_k -\hat{\theta}_k = \pm \sqrt{\mu(\vec{J}^{-1})_{kk}}
\end{equation}
This observation concludes the proof. 

\textbf{Lemma 2.} Let $\vec{\theta}$ have a uniform distribution on $B$, and let $C \subset B$ (here $B$ and $C$ can be arbitrary sets). Then $\vec{\theta} |\vec{\theta}\in C$ (i.e. the points of the distribution of $\vec{\theta}$ that fall in $C$) has a uniform distribution on $C$. \\
\noindent \textit{Proof.} $\vec{\theta}$ has a uniform distribution on $B$ if and only if, for any subset $C$ of $B$, 
\begin{equation}
    {\rm prob}(\vec{\theta} \in C)=\frac{V(C)}{V(B)}
\end{equation}
To show that $\vec{\theta} | \vec{\theta}\in C$ has a uniform distribution on $C$, we need to prove that for any subset $S$ of $C$ we  have:
\begin{equation}
    {\rm prob}(\vec{\theta}\in S | \vec{\theta} \in C) = \frac{V(S)}{V(C)}
\end{equation}
However:
\begin{align}
    {\rm prob}(\vec{\theta}\in S | \vec{\theta} \in C)&=\frac{{\rm prob}(\vec{\theta}\in S \cap \vec{\theta} \in C)}{{\rm prob}(\vec{\theta} \in C)} \nonumber \\
    &=\frac{{\rm prob}(\vec{\theta} \in S )}{{\rm prob}(\vec{\theta}\in C)}=\frac{V(S)}{V(B)}/ \frac{V(C)}{V(B)}=\frac{V(S)}{V(C)}
\end{align}
and the proof is finished. \\
\noindent \textit{Remark:} the above results are known but we  have included proofs of them to make this paper as self-contained as possible. 

\noindent  \textbf{The acceptance and rejection algorithm:} 
Let $f(\vec{\theta}), \vec{\theta} \in \mathcal{R}^{d}$, be a  pdf from which we want to sample, and let $g(\vec{\theta})$ be another pdf from which samples can be generated more easily than from $f(\vec{\theta})$, and which satisfies
\begin{equation}
    \frac{f(\vec{\theta})}{cg(\vec{\theta})} \leq 1 \ {\rm for \ a \ constant \ } c\geq 1 \ {\rm and } \ \forall \vec{\theta} \ (g(\vec{\theta}) \ \neq 0) 
\end{equation}
Then the samples $\vec{\theta}$ generated by the following two-step algorithm have the desired pdf  $f(\vec{\theta})$ (see, e.g., \cite{liu2001monte}):
\begin{itemize}
    \item Generate $\vec{\theta}$ from $g(\vec{\theta})$ and $u$ from a uniform distribution on $\left(0, 1 \right)$.
    \item If $u \leq f(\vec{\theta}) / (cg(\vec{\theta}))$ then accept $\vec{\theta}$, else reject $\vec{\theta}$ and repeat the two steps.
\end{itemize}

The manners in which $\lbrace  \vec{\theta}^m \rbrace$ were generated in both UE and UEG (as well as GE) are special cases of the above algorithm. In the case of UE,
\begin{equation}
    f(\vec{\theta})=\left\{\begin{aligned}
        1/V(C) \quad  &{\rm  if} \quad \vec{\theta} \in C \\
        \! 0  \quad & {\rm else} \\ 
\end{aligned}
\right.
\end{equation}
\begin{equation}
    g(\vec{\theta})=\left\{\begin{aligned}
        1/V(B) \quad  &{\rm  if} \quad \vec{\theta} \in B\\
        \! 0  \quad & {\rm else} \\ 
\end{aligned}
\right.
\end{equation}
and $f(\vec{\theta}) / (cg(\vec{\theta}))=1$ for $c=V(B)/ V(C)$, $\forall \vec{\theta} \in C$ (and =0 for $\vec{\theta}\in B \setminus C$). In the case of UEG (and GE), 
\begin{align}
    f(\vec{\theta}) = \eqref{eq:23} \nonumber \\
    g(\vec{\theta}) = \eqref{eq:20} \nonumber
\end{align}
and we also have $f(\vec{\theta})/(cg(\vec{\theta}))=1$ for $c=1/ \rho$ and $\forall \vec{\theta} \in C$ (and =0 for $\vec{\theta} \notin C$). In both cases all $\vec{\theta}$  vectors that fall in $C$ are accepted, and the other $\vec{\theta}$ vectors that are not in $C$ are rejected. (we have omitted the sub-index $n$ throughout this discussion to simplify the notation.)

\noindent  \textbf{Variance analysis of Monte-Carlo sampling:} To simplify the notation we omit the index $n$ of different variables and write $p(\vec{y}|\mathcal{M},\vec{\theta})$ and $p(\vec{\theta} | \mathcal{M})$ as $p(\vec{y}|\vec{\theta})$ and $p(\vec{\theta})$. We also define (see \eqref{eq:7} and \eqref{eq:17}): 
\begin{align}
    p&=\int_{D}p(\vec{y}|\vec{\theta})p(\vec{\theta})d\vec{\theta} \quad D\subset \mathcal{R}^d \label{eq:44} \\
    \hat{p}&=\frac{1}{M} \sum_{m=1}^M p(\vec{y}|\vec{\theta}^m) \quad \lbrace \vec{\theta}^m 
    \rbrace={\rm independently \ drawn \ from}\ p(\vec{\theta}). \nonumber
\end{align}
First we show that $\hat{p}$ is an \textit{unbiased estimate} of $p$:
\begin{equation}
    {\rm E}\left[ \hat{p}\right]=\frac{1}{M}\sum_{m=1}^M {\rm E}\left[p(\vec{y}|\vec{\theta^m}) \right]
    =\frac{1}{M}\sum_{m=1}^M\int_{D} p(\vec{y}|\vec{\theta})p(\vec{\theta})d\vec{\theta}=p.
\end{equation}
Next we derive a \textit{formula for the variance} of $\hat{p}$:
\begin{align}
    {\rm var}(\hat{p})=&{\rm E}\left[\hat{p}-p\right]^2={\rm E}\left\{\frac{1}{M}\sum_{m=1}^M\left[p(\vec{y}|\vec{\theta}^m)-p\right] \right\}^2=\frac{1}{M^2} \sum_{m=1}^M\sum_{\bar{m}=1}^M{\rm E}\left[p(\vec{y}|\vec{\theta}^m)-p \right]\left[ p(\vec{y}|\vec{\theta}^{\bar{m}})-p\right] \nonumber\\
    &=\frac{1}{M^2} \sum_{m=1}^M {\rm E}\left[p(\vec{y}|\vec{\theta}^m)-p \right]^2 =\frac{1}{M^2} \sum_{m=1}^M \left[ {\rm E}\left[ p(\vec{y}|\vec{\theta}^m)\right]^2-p^2 \right] \nonumber \\
    &=\frac{1}{M}\left[ \int_{D}p^2(\vec{y}|\vec{\theta})p(\vec{\theta})d\vec{\theta}-p^2 \right] \label{eq:46}
\end{align}
where we used the fact that the random variables $\lbrace p(\vec{y}|\vec{\theta}^m) -p\rbrace$ are independent and have zero mean.
Note from \eqref{eq:46} that
\begin{align}
    M{\rm var}(\hat{p})=\int_{D}\left[ p\sqrt{p(\vec{\theta})} -p(\vec{y}|\vec{\theta})\sqrt{p(\vec{\theta})} \right]^2d\vec{\theta}. \label{eq:48}
\end{align}
The above equation shows that the ideal case of ${\rm var}(\hat{p})=0$ is achieved if and only if
\begin{equation}
     p(\vec{y}|\vec{\theta})={\rm const}.
\end{equation}
This observation lies at the basis of the importance sampling method described in the Section titled ``UEG''. 

From \eqref{eq:46} we also see that:
\begin{equation}
    {\rm var}(\hat{p})\leq \frac{1}{M} \int_{D} p^2(\vec{y}|\vec{\theta})p(\vec{\theta})d\vec{\theta} \leq \frac{p^2(\vec{y}|\hat{\vec{\theta}})}{M} \label{eq:50}
\end{equation}
where $\hat{\vec{\theta}}$ is the MLE of $\vec{\theta}$. The numerator in \eqref{eq:50} is typically much smaller than one (especially for large values of $N$) and consequently ${\rm var}(\hat{p})$ may take on quite acceptable values in many cases even for reasonably small values of $M$ such as $M=1000$. However when we compare model structures with one another, the differences between the corresponding values of $p$ in \eqref{eq:44}  can sometimes be rather small (especially for large values of $d$) and in such cases much larger values of $M$ than $M=1000$ may be needed to reduce the estimation errors in $\hat{p}$ under the level of the aforementioned differences.

It follows from \eqref{eq:50} that a direct way of reducing the var is by increasing $M$ (as expected). A theoretically more interesting way called "stratified sampling", which in particular applies to UB, is described in the next subsection.

\noindent\textbf{Stratified sampling:} Split $D$ in $K$ non-overlapping sub-sets $\lbrace D_k \rbrace$:
\begin{equation}
    \cup_{k=1}^K D_k=D
\end{equation}
Let $V$ and $V_k$ denote the volumes of $D$ and $D_k$, respectively, and let $\rho_k=V_k/ V$. Note that:
\begin{equation}
    \sum_{k=1}^K \rho_k=1
\end{equation}
Finally, let $p_k(\vec{\theta})$ denote the uniform distribution on $D_k$:
\begin{equation}
    p_k(\vec{\theta})=\left\{\begin{aligned}
        1/V_k \quad  &{\rm  for} \quad \vec{\theta} \in D_k \\
        \! 0  \quad & {\rm else} \\ 
\end{aligned}
\right.
\end{equation}
We will use the same notation as in the previous subsection of the Appendix, and make use of the variance result proved there.
\par The quantity we want to estimate is $p$, which can be rewritten as:
\begin{equation}
    p=\int_{D}p(\vec{y}| \vec{\theta})p(\vec{\theta})d\vec{\theta}=\sum_{k=1}^K \int_{D_k} p(\vec{y}|\vec{\theta})p(\vec{\theta})d\vec{\theta}=\sum_{k=1}^K \rho_kp_k
\end{equation}
where $p(\vec{\theta})$ denotes the uniform pdf on $D$, and
\begin{equation}
    p_k=\int_{D_k} p(\vec{y}|\vec{\theta})p_k(\vec{\theta})d\vec{\theta}
\end{equation}
The stratified sampling method estimates $p_k \ ({\rm for \ }k=1,\dots,K)$ using $\rho_kM$ samples (assuming, for simplicity, that $\rho_kM$ is an integer) independently drawn from $p_k(\vec{\theta})$, and  then combines the so-obtained estimates $\lbrace \hat{p}_k \rbrace$ into an estimate of $p$:
\begin{equation}
    \hat{p}_{S}=\sum_{k=1}^K \rho_k\hat{p}_k
\end{equation}
Interestingly, the variance of $\hat{p}_S$ is always smaller than the variance of the non-stratified estimate that was analysed in the previous subsection (see \eqref{eq:46} there):
\begin{equation}
    {\rm var}(\hat{p}_S)\leq \frac{1}{M} \left[ \int_{D} p^2(\vec{y}|\vec{\theta})p(\vec{\theta})d\vec{\theta} -p^2 \right] \label{eq:57}
\end{equation}
To prove the above inequality, first note that the bias of $\hat{p}_k$ is zero and its variance is given by the following formula (similar to \eqref{eq:46}):
\begin{equation}
    {\rm var}(\hat{p}_k)= \frac{1}{\rho_kM}\left[\int_{D_k} p^2(\vec{y}|\vec{\theta})p_k(\vec{\theta})d\vec{\theta}-p_k^2 \right]
\end{equation}
Because 
\begin{equation}
    \hat{p}_S-p=\sum_{k=1}^K\rho_k(\hat{p}_k-p_k)
\end{equation}
where the terms are independent of each other, one can readily check that $\hat{p}_S$ is an unbiased estimate of $p$ with the  following variance:
\begin{equation}
    {\rm var}(\hat{p}_S)=\sum_{k=1}^K \rho_k^2 {\rm var}(\hat{p}_k) =\frac{1}{M} \sum_{k=1}^K \rho_k\left[\int_{D_k} p^2(\vec{y}|\vec{\theta})p_k(\vec{\theta})d\vec{\theta} -p_k^2 \right] \label{eq:60}
\end{equation}
The first term in \eqref{eq:60}, namely
\begin{equation}
    \frac{1}{M} \sum_{k=1}^K \rho_k \int_{D_k} p^2(\vec{y}|\vec{\theta})p_k(\vec{\theta})d\vec{\theta}=\frac{1}{M}\int_{D}p(\vec{y}|\vec{\theta})p(\vec{\theta})d\vec{\theta}
\end{equation}
is the same as the corresponding term in \eqref{eq:46}. However, the second term in \eqref{eq:60} is smaller than its counterpart in \eqref{eq:46}, which is a consequence of the Cauchy-Schwarz inequality:
\begin{equation}
    p^2=\left( \sum_{k=1}^K \rho_kp_k \right)^2=\left( \sum_{k=1}^K \sqrt{\rho_k} \cdot \sqrt{\rho_k}p_k\right)^2 \leq \underbrace{\left(\sum_{k=1}^K \rho_k \right)}_{=1} \left(\sum_{k=1}^K \rho_kp^2_k \right)=\sum_{k=1}^K \rho_kp^2_k \label{eq:62}
\end{equation}
This observation concludes the proof of \eqref{eq:57}.

Note that the equality in \eqref{eq:62}, and hence in \eqref{eq:57}, holds if $\sqrt{\rho_k}={\rm const}\cdot \sqrt{\rho_k}p_k$, i.e. $p_k={\rm const}$ (for $k=1,\dots,K$). However, in applications $p(\vec{y}|\vec{\theta})$ often has dominant peaks and therefore $\lbrace p_k \rbrace$ is far from being a constant sequences. In such cases the stratified approach can  improve the accuracy of the Monte-Carlo sampling in a meaningful way.

We have compared the results obtained with UB and those provided by stratified sampling in a number of cases.  The latter method was always better but the gain (i.e. the increase in the probability of correct selection) was relatively modest. More concretely, we divided each axis of the box $B$ in $L$ segments which yielded a number of sub-boxes  that increased exponentially in $d$, namely $L^d$ sub-boxes. In our experiments the accuracy of  stratified sampling increased with $L$ so we chose $L$ so that $L^d$, for the maximum value of $d$, was smaller than but as close to $M$ as possible. The accuracy gain achieved by stratified sampling was comparable to but slightly smaller than that we obtained using UB with $10M$. However in the cases in which increasing $M$ several times is not possible due to the incurred increase of computation time, stratified sampling is a possible approach to improving the UB accuracy at basically no increase in computing time.

\bibliographystyle{IEEEtran} 
\normalsize
\bibliography{main}

\begin{thebibliography}{10}
\providecommand{\url}[1]{#1}
\csname url@samestyle\endcsname
\providecommand{\newblock}{\relax}
\providecommand{\bibinfo}[2]{#2}
\providecommand{\BIBentrySTDinterwordspacing}{\spaceskip=0pt\relax}
\providecommand{\BIBentryALTinterwordstretchfactor}{4}
\providecommand{\BIBentryALTinterwordspacing}{\spaceskip=\fontdimen2\font plus
\BIBentryALTinterwordstretchfactor\fontdimen3\font minus
  \fontdimen4\font\relax}
\providecommand{\BIBforeignlanguage}[2]{{%
\expandafter\ifx\csname l@#1\endcsname\relax
\typeout{** WARNING: IEEEtran.bst: No hyphenation pattern has been}%
\typeout{** loaded for the language `#1'. Using the pattern for}%
\typeout{** the default language instead.}%
\else
\language=\csname l@#1\endcsname
\fi
#2}}
\providecommand{\BIBdecl}{\relax}
\BIBdecl

\bibitem{kay1993fundamentals}
S.~M. Kay, \emph{Fundamentals of statistical signal processing: Detection
  theory}.\hskip 1em plus 0.5em minus 0.4em\relax Prentice-Hall Englewood
  Cliffs, NJ, 1993, vol.~II.

\bibitem{stoica2004model}
P.~Stoica and Y.~Sel{\'e}n, ``Model-order selection: {A} review of information
  criterion rules,'' \emph{IEEE Signal Processing Magazine}, vol.~21, no.~4,
  pp. 36--47, 2004.

\bibitem{stoica2004cross}
------, ``Cross-validation rules for order estimation,'' \emph{Digital Signal
  Processing}, vol.~14, no.~4, pp. 355--371, 2004.

\bibitem{stoica2012proper}
P.~Stoica and P.~Babu, ``On the proper forms of {BIC} for model order
  selection,'' \emph{IEEE Transactions on Signal Processing}, vol.~60, no.~9,
  pp. 4956--4961, 2012.

\bibitem{llorente2020marginal}
F.~Llorente, L.~Martino, D.~Delgado, and J.~Lopez-Santiago, ``Marginal
  likelihood computation for model selection and hypothesis testing: {An
  extensive review},'' \emph{arXiv preprint arXiv:2005.08334}, 2020.

\bibitem{soderstrom1989system}
T.~S{\"o}derstr{\"o}m and P.~Stoica, \emph{System identification}.\hskip 1em
  plus 0.5em minus 0.4em\relax London, England: Prentice Hall International,
  1989.

\bibitem{kay1998fundamentals}
S.~M. Kay, \emph{Fundamentals of statistical signal processing: {Estimation
  theory}}.\hskip 1em plus 0.5em minus 0.4em\relax Prentice-Hall Englewood
  Cliffs, NJ, 1998, vol.~I.

\bibitem{white1996estimation}
H.~White, \emph{Estimation, inference and specification analysis}.\hskip 1em
  plus 0.5em minus 0.4em\relax Cambridge university press, 1996, no.~22.

\bibitem{mcbook}
\BIBentryALTinterwordspacing
A.~B. Owen, \emph{Monte Carlo theory, methods and examples}, 2013. [Online].
  Available: \url{http://www-stat.stanford.edu/~owen/mc}
\BIBentrySTDinterwordspacing

\bibitem{robert2013monte}
C.~Robert and G.~Casella, \emph{Monte {Carlo} statistical methods}.\hskip 1em
  plus 0.5em minus 0.4em\relax Springer Science \& Business Media, 2013.

\bibitem{liu2001monte}
J.~S. Liu, \emph{Monte {Carlo} strategies in scientific computing}.\hskip 1em
  plus 0.5em minus 0.4em\relax Springer, 2001, vol.~10.

\end{thebibliography}

\end{document}